\input harvmac
\def\psqr#1#2{{\vcenter{\vbox{\hrule height.#2pt
	\hbox{\vrule width.#2pt height#1pt \kern#1pt
	\vrule width.#2pt}
	\hrule height.#2pt \hrule height.#2pt
	\hbox{\vrule width.#2pt height#1pt \kern#1pt
	\vrule width.#2pt}
	\hrule height.#2pt}}}}
\def\sqr#1#2{{\vcenter{\vbox{\hrule height.#2pt
	\hbox{\vrule width.#2pt height#1pt \kern#1pt
	\vrule width.#2pt}
	\hrule height.#2pt}}}}

\def\doub{\mathchoice\psqr65\psqr65\psqr{2.1}3\psqr{1.5}3}

\hfill hep-th/9812015
\vskip 1.5in
\centerline{\bf Molien Function for Duality}
\bigskip\bigskip
\centerline{Philippe Pouliot}
\bigskip
\centerline{Department of Physics}
\centerline{University of California}
\centerline{Santa Barbara, CA 93106}
\vskip 1in
\centerline{Abstract}
\bigskip
The Molien function counts the number of independent group 
invariants of a representation. For chiral superfields, it is
invariant under duality by construction. 
We illustrate how it calculates the
spectrum of supersymmetric gauge theories.
\vfill\eject

\nref\intsei{K. Intriligator and N. Seiberg, Lectures on  
Supersymmetric Gauge Theories and Electro-Magnetic Duality,
Nucl. Phys. suppl. 45BC (1996) 1, hep-th/9509066.}
\nref\sattinger{D. Sattinger and O. Weaver, Lie Groups and Algebras
with Applications to Physics, Geometry and Mechanics, 1986
Springer-Verlag.}
\nref\forger{M. Forger, Invariant Polynomials and Molien Functions,
J. Math. Phys. 39 (1998) 1107.}
\nref\cummins{L. Begin, C. Cummins and P. Mathieu, Generating Functions
for Tensor Products, hep-th/9811113.}
\nref\iss{K. Intriligator, N. Seiberg and S. Shenker, 
Proposal for a Simple Model of Dynamical SUSY
Breaking, Phys. Lett. B 342 (1995) 152, hep-th/9410203.}

{\bf 1. Introduction}
\bigskip
It is quite remarkable that certain four dimensional gauge theories
can be solved exactly. The examples that have been solved so far
(\intsei\ for a review) are quite special: they have lots of 
symmetries. 

A generic theory does not have so many symmetries, so here I
introduce a tool which I hope will be useful to the study of more
general theories. To be concrete, I will consider only supersymmetric
theories.

\bigskip
{\bf 2. The Molien function}
\bigskip
Consider a supersymmetric gauge theory with chiral superfields
transforming as a representation $R$ of a group $G$. 
I make no restrictions on
$R$: it can be reducible, and also contain singlets (mesons).
Similarly, $G$ can be a product of groups, or it can be the identity,
for a confining theory.

The Molien generating function for the representation $R$ is
$$
M(z) = \sum^{\infty}_{k=0} c_k z^k
$$
where $c_k$ is the number of independent 
group invariant polynomials of order $k$. 
It is a holomorphic function. 

It turns out that there is a nice way to write down $M(z)$
(see \sattinger\ p. 204 for an easy proof):
$$
M(z) = \int {d\mu(g)\over \det (1 - z R(g))}.
$$ 
The idea of the proof is that one can diagonalize 
the unitary representation $R$ for any fixed
group element $g$; then integration ($\int d\mu(g)$) 
over the whole group picks out only the singlets in the tensor
products $R^{\otimes k}$. 

The function $M$ can
be evaluated more explicitly (see the nice paper by Forger \forger\
for much useful and readable complementary details.)
$$
M(z)= {1\over |W|}\int \cdots \int {dw_1\over 2\pi i w_1}\cdots 
{dw_l\over 2\pi i w_l} {\Pi_\alpha (1 - w^{h(\alpha)})\over
\Pi_\lambda (1- z w^{h(\lambda)})}.
$$

$\bullet$ $|W|$ is the number of elements in the Weyl group

$\bullet$ $l$ is the rank of the group;

$\bullet$ the products are over all the roots $\alpha$ of the group 
and over all the weights $\lambda$ of the representation $R$;

$\bullet$ and finally, the concise notation $w^{h(\alpha)}$ means
$w_1^{h(\alpha_1)}\cdots w_l^{h(\alpha_l)}$, where $h(\alpha_i)$
is the eigenvalue of the root $\alpha$ under the Cartan generator
$H_i$.

Another representation of the Molien function coefficients
is given in terms of an index:
$$
c_k = {1\over |W|} \sum_{\tilde \lambda} i(\tilde\lambda)
m_k(\tilde\lambda)
$$
with the $\tilde\lambda$ denoting the extended weights and $m_k$
the $k$-extended multiplicities. This terminology is defined in
\forger. According to \forger, 
the index might be more efficient for explicit calculations.

\bigskip
{\bf 3. Duality\footnote{$^1$}{I wish to thank O. Aharony and A.
Schwimmer for inquiries that led to a clarification of this
section.}}
\bigskip

For a ${\cal N}=1$ supersymmetric 
gauge theory with a vanishing superpotential, the Molien function
calculated for the gauge group of the theory
encodes much information about the low-energy spectrum. It
calculates how many gauge invariant independent 
chiral (holomorphic) operators
there are of a given degree in the number of elementary fields.
In other words, it contains much about the structure of 
the chiral ring. 

Consider now a dual ``magnetic'' description to this theory. Since by
assumption it has the same low-energy spectrum, there must be a
 way, expected to be complicated (i.e. {\bf not} a Molien function!), 
to calculate the Molien function of the ``electric''
theory in terms of the data of the magnetic theory. In this sense
the Molien function is duality invariant.

Since it is a holomorphic function, one would
hope that the powerful tools of complex analysis can be useful
to study its properties. 
This, however, is highly speculative, as well as the remaining
of this paragraph. 
Even if it is not enough to fully characterize
a supersymmetric conformally invariant gauge theory, 
the Molien function could
be ``the'' characteristic function of ${\cal N} = 1$ duality
\footnote{$^2$}{
(However, there is no claim of uniqueness here: with the meaning
of duality invariance above, any formal function of variables
$z_\alpha$ (one variable for each chiral invariant operator $O_\alpha$)
will be duality invariant, as long as the constraints among the
operators are suitably implemented by constraints among the
$z_\alpha$, a feat that the Molien function accomplishes naturally. 
See section 4.)}.
It is definitely 
interesting because it does not rely on global symmetries. 
For a generic
theory, global symmetries are small and the constraints one
can get from them have a limited power: it is well known that  
satisfying the 't Hooft anomaly
matching conditions is not enough. In string theory, there are no
global symmetries anyway. Perhaps dynamical properties
of the low-energy theory can be inferred from the chiral spectrum.
This would come about by making the following statement more precise: 
the Molien function of a confining theory is simple, while
the Molien function
 of a gauge theory which is not asymptotically free is
complicated (it has high order syzygies among its invariants.)

If the theory has a non-zero superpotential, extra constraints are
introduced among the invariants. The definition of the Molien function
stays the same (namely $M(z) = \sum^{\infty}_{k=0} c_k z^k$
where $c_k$ is the number of independent 
group invariant polynomials of order $k$),
but the integral representation ($\int {d\mu(g)\over 
\det (1 - z R(g))}$) should be generalized to include the effect of the superpotential
(I don't know how to write it down).
\bigskip
{\bf 4. Generalized Molien Function}
\bigskip
It would be nice to have an explicit way to construct the
Molien function of the electric from the magnetic theory. 
One might hope that there is a generalization of the Molien function,
$\tilde M$, which is such that calculating $\tilde M$ in the electric
and in the magnetic theory would give the same result. I do not
know if this is possible.
 
As a step in this direction, it is
convenient to define a generalized Molien function, still assuming
that the superpotential is zero, by choosing a global $U(1)$ charge,
under which the elementary fields in irreducible
representation matrices $R_i$ transform with charges $q_i$,
$i=1,\ldots,n$ (which can all
be taken to be integers by suitable rescaling). With this,
the generalized Molien function
$$
M_{\{q_i\} } = \int d\mu(g)/ \det 
\pmatrix{1 - z^{q_1} R_1(g)) \cr
& 1 - z^{q_2} R_2(g)) \cr
 & & \ddots \cr
& & & 1 - z^{q_n} R_n(g)) \cr}
$$
has the property that the invariant operators are counted with the
same power of $k$ as coefficients of $z^k$ in the electric and in the
magnetic theories. The coefficients will still 
disagree of course because
the constraints from the superpotentials have not been included.

\bigskip
{\bf 5. Illustration}
\bigskip

Aside from the duality application, 
the Molien function provides a technique to grind
out the spectrum of a theory, along with plethysms, branching rules
and other counting arguments.
I will illustrate some of the uses with the simplest examples.
Start with ${\cal N}=1$ supersymmetric $SU(2)$ gauge theory with
one flavor of fundamentals $Q_i$ (two doublets)~\intsei.

Evaluating $M(z)$ with the integral representation readily gives
$$
M = {1 \over 1- z^2} = { 1 + z^2 + z^4 + \cdots}.
$$
This generating function is characteristic of a freely generated
ring with one invariant: there's one polynomial of order 2, namely 
$Q_1Q_2$, and one of order 4, $(Q_1Q_2)^2$, and so on.

With 4 doublets, 
$$
M(z) ={ 1 - z^4\over (1-z^2)^6} = 1+ 6 z^2 + 20 z^4 + 50 z^6 +\cdots.
$$
The coefficient $6$ indicates that the ring is generated by the
invariants $V_{ij} = Q_iQ_j$. At order $z^4$, we learn that the 
$V_{ij}$ are not independent, but there is one constraint among them,
the famed $\rm{pf}\ V = \Lambda^4$. Studying the following coefficients
shows that there are no more constraints.

With 6 doublets,
$$
M(z) = {1+ 6 z^2 + 6 z^4 + z^6\over (1-z^2)^9} = 1+ 15 z^2 +
(120 - 15)  z^4 + (680 - 189 - 1) z^6 +\cdots.
$$
This is already more complicated. There are 15 invariants $V_{ij}$,
and $15$ constraints (syzygies)
$\epsilon^{ijklmn}V_{kl}V_{mn}$, but there
are constraints amongst the constraints and so on.

More generally, with $d$ doublets, 
$$
M_{d} = \sum_{k=0}^\infty \dim \left( \ \doub\doub\cdots \doub\ 
\right) z^k
$$
where the tensor under the $SU(2d)$ symmetry has $k$ horizontal boxes. 
\bigskip
{\bf 6. New Example}

\bigskip
As the rank of the group increases, the formulas for the Molien 
function become rapidly cumbersome to evaluate. For the integral
representation, one is faced with high order poles to be evaluated by
the residue theorem. A trick is to settle for less than the full
generating function, and get only 
 the first few $c_k$: one takes
derivatives with respect to $z$, and then set $z=0$, before 
evaluating the residues at the $w_i$, for which the
poles are now automatically
all at $w_i = 0$. To go beyond that,
perhaps one could reexpress these integrals
using Littlewood's Schur functions, or use the index formula
of \forger. Another possibility to evaluate the Molien function more
effectively is to use the MacMahon algorithm \cummins 
\footnote{$^3$}{I thank C. Cummins for pointing out 
the usefulness of \cummins\ in this
respect.}.

\bigskip
Here for simplicity,
 I will only calculate the spectrum of the supersymmetric $SU(2)$ 
gauge theories with one matter field in the 4-dimensional 
representation $S$ and $2k$ doublet fields $Q_i$.
\bigskip
$\bullet$ $k = 0$.
This theory was studied in \iss. There is just one invariant, quartic,
so $M = 1/(1-z^4)$. 
\bigskip
When $k>0$, the theories are not asymptotically free. That does not
make them uninteresting, because they can still be the free duals
of strongly coupled theories.

$\bullet$ $k = 1$
$$
M = {1-z^2+5 z^4 - z^6 + z^8\over (1-z^4)^3 (1- z^2)^2 }
$$
At this stage, we see the invariants $Q^2$, $SQ^3$,
$S^2Q^2$, $S^3Q^3$ and $S^4$. The invariants $SQ^3$,
$S^2Q^2$, $S^3Q^3$ are fully symmetric in their flavor indices.
They generate the full ring, but they
are not independent. Checking this result for $k>1$, we see that these
invariants still form a full set, but there are more constraints.
\bigskip
$\bullet$ $k=2$

$$
M = {1 + 2z^2 + 28z^4 + 23z^6 + 73 z^8 + 23z^{10} + 28z^{12} + 
      2z^{14} + z^{16}\over (1 - z^4)^5(1 - z^2)^4}.
$$

$\bullet$ $k=3$

$
M = {1 + 9z^2 + 101z^4 + 319z^6 + 1020z^8 + 1475z^{10} + 
      2091z^{12} + 1475z^{14} + 1020z^{16} + 319z^{18} + 
      101z^{20} + 9z^{22} + z^{24}\over (1- z^4)^7(1 - z^2)^6}.
$

\bigskip
\centerline{Acknowledments}
\bigskip
This work is supported by NSF Grant PHY97-22022. I wish to thank
O. Aharony, C. Cummins and A. Schwimmer for useful correspondence.
\bigskip
\listrefs 
\bye